\RequirePackage{lineno}
\documentclass[aps,reprint,notitlepage,nofootinbib,amssymb, preprintnumbers,amsmath,groupedaddress,superscriptaddress,showkeys]{revtex4-1}
\pdfoutput=1
\usepackage{graphicx} 
\usepackage{hyperref} 
\usepackage{cases}
\usepackage{bbm}
\usepackage{color}
\usepackage{dsfont}
\usepackage{enumitem}
\usepackage{relsize}
\usepackage{mathtools}
\usepackage{lineno}
\usepackage{latexsym}
\usepackage{graphicx}
\usepackage[normalem]{ulem}
\newcommand{\e}{\mathrm{e}}
\newcommand{\Prob}{\ensuremath{\mbox{\small Prob}}}

\begin{document}

\title{Effective bandwidth of non-Markovian packet traffic}

\author{Massimo Cavallaro}
\email[]{m.cavallaro@warwick.ac.uk}
\affiliation{School of Life Sciences and Department of Statistics, University of Warwick, Coventry, CV4 7AL, UK}
\author{Rosemary J. Harris}
\email[]{rosemary.harris@qmul.ac.uk}
\affiliation{School of Mathematical Sciences, Queen Mary University of London, Mile End Road, London, E1 4NS, UK}
\begin{abstract}

We demonstrate the application of recent advances in statistical mechanics to
a problem in telecommunication engineering: the assessment
of the quality of a communication channel in terms of rare and extreme events.
In particular, we discuss non-Markovian models for telecommunication traffic in
continuous time  and deploy the ``cloning'' procedure of non-equilibrium
statistical mechanics to efficiently compute  their effective bandwidths.  The
cloning method allows us to evaluate the performance of a traffic protocol
even in the absence of analytical results, which are often hard to obtain when
the dynamics are non-Markovian.

\end{abstract}



\maketitle

{\tableofcontents}

\section{Introduction}
\label{sec:Introduction}

Natural systems made of many coupled components, ranging from ideal gases to
living organisms and their communities, have long been of interest to
scientists. Recently, by contrast, some of the most  studied complex systems
have been man-made, for instance telecommunication networks, transport
infrastructures, and financial markets. The methods used to approach these
technological systems are similar to those used in natural sciences. Indeed,
at a sensible level of detail, the system properties appear as random
variables, and the scientific effort is directed towards the quantification of
such randomness, as well as of its effects. More specifically, in
telecommunication engineering, we are interested in relating the elementary
(``microscopic'') description of a telecommunication network in terms of
packets and servers, to a perceivable (``macroscopic'') quantity, such as the
service that is  effectively available to the final user. Such a programme
looks very similar to the one that led to the the development of statistical mechanics, where
macroscopic observables (such as density and energy) are defined based on
microscopic modelling.

Studies comparing communication networks and many-body physical systems are
well-represented in the literature~\cite{DeMartino2009,DeMartino2009a,
Chernyak2010} and it is now well understood that the analogies are based on
the underlying mathematics of stochastic processes and large deviation
theory~\cite{Touchette2009}. This paper is built around the less well-known
notion of \textit{effective bandwidth} (EB), which has been introduced in the 90s
to weight the effects of large deviation events in
resource allocation~\cite{Kelly1996},
(see also the more recent reference books~\cite{Gautam2012,Srikant2013,Kelly2014,Larsson2014}). It is worth
noting that the efficient numerical evaluation of rare events is central
within such a context, as in the real world a suitable amount of resources
must be allocated even in the absence of analytical solutions for the
probability of  potentially disruptive rare events~\cite{Heidelberger1995,
Reijsbergen2012, Chang1994}. To do so, we choose a tool from non-equilibrium
statistical mechanics, viz., the ``cloning'' method~\cite{Giardina2011}. We
are here particularly interested in systems with non-Markovian dynamics
(relevant to real-world applications as well as statistical mechanics
advances) and so we exploit a recent implementation of cloning for
non-Markovian processes~\cite{Cavallaro2016}.

The text is organised as follows.
In section \ref{sec:Net_load} some basic concepts from queuing
theory are introduced to set the stage for the following sections.
In section~\ref{sec:Effective_bandwidth} we present, motivate, and review the notion of
EB, which is connected to a large deviation analysis of the net load.
In sections \ref{sec:fluid_process}, \ref{sec:MMPP}, and \ref{sec:Numerical_results_for_non-markovian_modulation}
we analyse Markovian and non-Markovian models of packet traffic modulated by an underlying stochastic process.
The cloning method is used to numerically compute their EBs.
In particular, while  sections \ref{sec:fluid_process} and \ref{sec:MMPP} deal with classic teletraffic toy models
whose EBs are known analytically,
in section~\ref{sec:Numerical_results_for_non-markovian_modulation}
we explore numerically the case of a modified non-Markovian two-phase process,
observing that the EB increases as the dispersion
of the phase lengths increases.
Significantly, this leads to the observation that the EB in the new model can be adjusted by tuning the traffic parameters,
whilst maintaining a fixed mean traffic rate.
We conclude with a discussion in section~\ref{sec:discussion}.

\section{Net load, bandwidth, and their bounds}
\label{sec:Net_load}

In a queuing network, we have a number of servers that exchange packets or
customers. When a collection of packets leaves a server to reach another one,
we say that a communication channel has been established between the two
servers. In such a channel, we denote the random amount of work brought by
customers passing through the channel during the interval $[0,t)$ as $X(t)$. This can
be described as the integral over $[0,t)$ of a process $\mathrm{x}(t)$, which
can be a ``point'' process (thought of as a series of random events describing
customers -- or particles -- feeding one of the servers), or a ``fluid''/``piece-wise deterministic'' process (supplied continuously and
deterministically between random times). We will not deal with a third
possibility where the instantaneous work is supplied continuously and
stochastically; however, many considerations in this paper arguably apply also
to such a case. Similarly, we consider the amount of work $Y(t)$ that the
receiving server is able to perform during the same amount of time, which is
the integral over $[0,t)$ of another point or fluid process $\mathrm{y}(t)$.
It is of central importance here that $X(t)$ and $Y(t)$ are assumed to be
time-extensive, so that they play the role of integrated currents in standard interacting
particle systems~\cite{Chernyak2010}. Indeed, such a setting is quite general.
For example, in the context of economics, $X(t)$ and $Y(t)$ can describe the
total production and demand, respectively.  It is also convenient to introduce
the quantity $W(t)=X(t)-Y(t)$, which we refer to as the \textit{net load}, and
its derivative $\mathrm{w}(t)$.

\begin{figure*}
  \centering
  \includegraphics[width=1\textwidth]{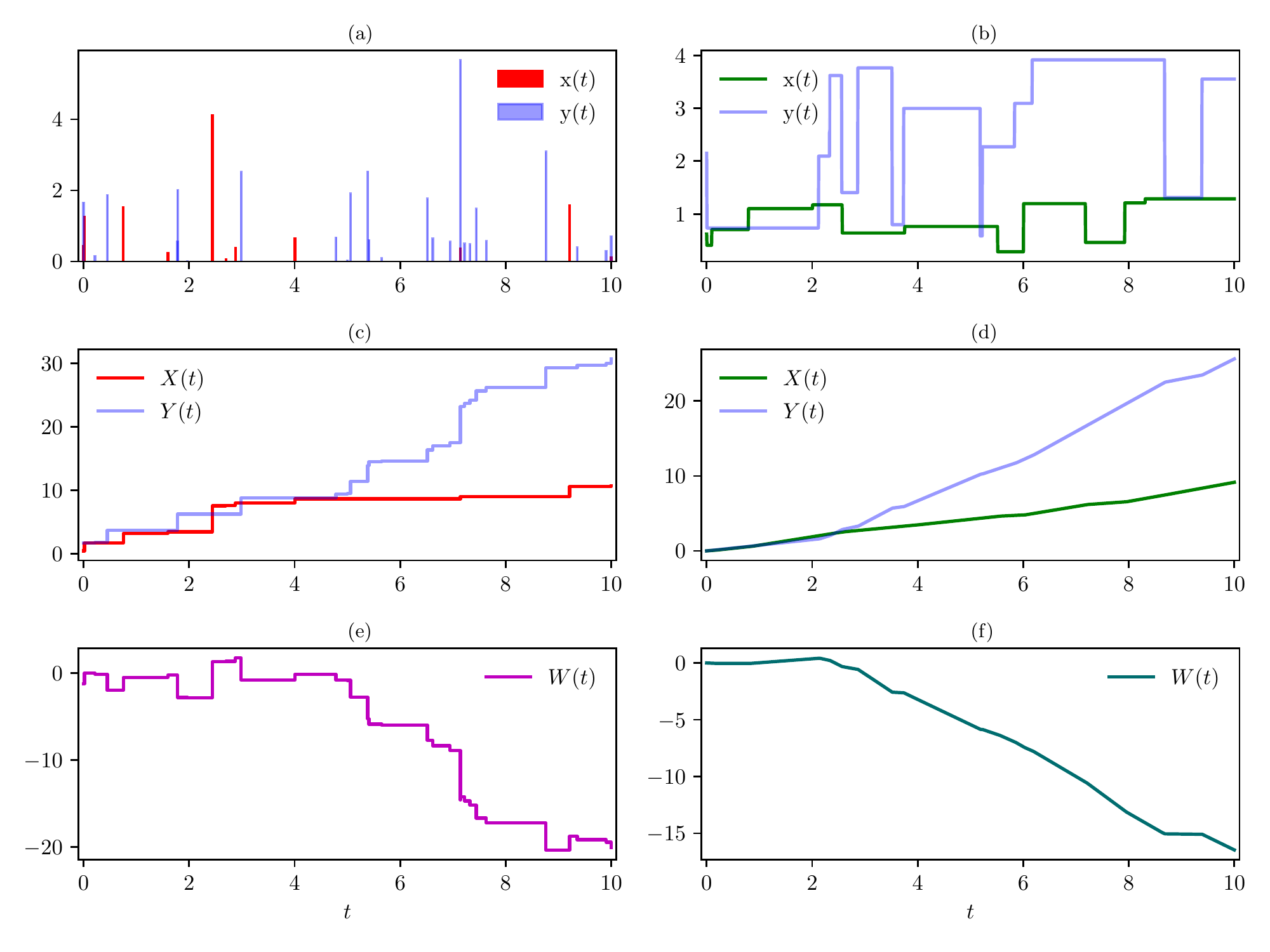}
  \caption{
  Left and right panels show example settings for point and fluid net loads, respectively.
  a) $\mathrm{x}(t)$ and $\mathrm{y}(t)$ represent discrete arrival and attempted service events, respectively.
  b) Both request $\mathrm{x}(t)$ and service $\mathrm{y}(t)$ are carried out continuously in time.
  The supply laws change at random time points, but are constant otherwise.
  c) and d) $X(t)$ is the total amount of work (e.g., customers to be served)
  requested during the interval $[0,t)$, while $Y(t)$ is the
  service that can be provided during the same amount of time. 
  e) and f) Resulting outcomes of the net-load processes $W(t)= X(t)-Y(t)$.
  \label{fig:net load_queue}
  }
\end{figure*}

Figure~\ref{fig:net load_queue} (left panels) shows an example of such a setting when both $X(t)$ and $Y(t)$ are point processes.
In this case, $W(t)$ has increments at the instants $(t_1,t_2,\ldots,t_n)$,
with $ 0 \le  t_1 \le \ldots \le t_n < t$, when discrete arrival or service events occur.
`
If $\mathrm{x}(t)$ or $\mathrm{y}(t)$ is fluid, then $W(t)$ increases or decreases deterministically between the random times $(t_1,t_2,\ldots,t_n)$.
This is illustrated
 in figure~\ref{fig:net load_queue} (right panels).
Obviously, it is possible to combine fluid and point processes.

In all the previous cases, the arrival, the service, and the net load can be regarded as driven by an underlying random process,
which influences or modulates their evolution in a point-wise or piecewise-deterministic fashion.
Conversely, the load itself has no influence on the modulating random process.
More specifically, in a point process, increments occur at the random times $t_i$, $i=1,2,\ldots,n$, and 
their frequency and magnitude $\theta_{t_i}$ can be thought of as being dictated by the modulating random process.
On the other hand, in a fluid process, the $t_i$s correspond to the transition times of
the underlying process and at the end of an interval between two consecutive transition times $t_i$ and $t_{i+1}$
the net load has increased by an amount that depends on $(t_{i+1} - t_i)$;
for simplicity, we assume that the fluid rate is constant between 
transition times and the load increase is $ (t_{i+1} - t_i) \theta_{t_{i}}$.
To maintain generality, we refer to all types of increment as~$\Theta_i$.

We require that the limits $\bar x = \lim_{t\to \infty} X(t)/t$ and $\bar y=\lim_{t\to \infty} Y(t)/t$ exist,
and we refer to the condition $\bar x < \bar y$ as \textit{stability}, which implies that $W(t)/t$ concentrates
around a negative value $\bar w = \bar x - \bar y$ as $t \to \infty$.
However, the service cannot be stored, and  it is possible 
that at any time there is a certain amount of work waiting to be done.
This loosely defines a non-negative random process that is referred to as
the queue length and is
denoted  by $Q(t)$. Its dynamics  are as follows.
For a point process $W(t)$, the work to be done at time
$t_n$ is the sum of the length of the queue at the previous event instant $t_{n-1}$ and the net-load
increment $W(t_{n})-W(t_{n-1})$ 
 during the same interval; however, when  $W(t_{n})-W(t_{n-1}) + Q(t_{n-1})<0$,
the work surplus is wasted, yielding a zero  queue length (instead of a negative one).
This can all be expressed  in the recursive relation
\begin{equation}
  Q(t_n) = \max\{0,W(t_{n})-W(t_{n-1}) + Q(t_{n-1}) \}.
  \label{eq:lindley}
\end{equation}
In the fluid case the queue dynamics is more subtle, but can be compactly described as
\begin{equation}
  Q(t) = \sup_{0 < \tau \le t} \int_{\tau}^{t}  \mathrm d u \, \mathrm{w}(u),
  \label{eq:queue}
\end{equation}
if the queue is empty at $t=0$, see, e.g., reference~\cite{Kelly2014}.
In this paper, we focus on net-load statistics but will also discuss
queue-length statistics.

A convenient simplification consists of assuming deterministic service.
When the service is continuously and deterministically provided at constant
rate $c$, the service offered during the time interval $[0,t)$ is $c t$,
and the net-load process  is 
\begin{equation}
  W(t) = X(t) - c t.
  \label{eq:net load_form}
\end{equation}
In telecommunications, the quantity $c$ is the data transfer rate of the communication channel,
which we refer to as the \textit{bandwidth}, and is a quantity that
can typically be controlled by the service provider.
In more general terms, $c$ is a reference deterministic rate at which a service can be provided.
In the following, unless explicitly stated,  we will deal with the EB of an arrival process $X(t)$ which feeds a net load of the form \eqref{eq:net load_form}.
However, the only condition required for the EB analysis of $X(t)$ is its extensivity, hence it is possible to perform a similar treatment for any observable, as long as it is time-extensive,
regardless of whether it is meant to describe the arrival, the service or the net load.

A very loose bound theorem on a generic random variable $X$ that takes
non-negative values with density function $f(u)$
can be derived from the knowledge of its expectation value, i.e.,
\begin{equation}
    \int_{0}^{\infty} \mathrm{d}u \, u f(u) \ge x \int_{x}^{\infty}\mathrm{d} u \, f(u) ,
    \label{eq:Markov_inequality_0}
\end{equation}
for $x \ge 0$. This  can be rewritten more conveniently as
\begin{equation}
    \Prob \{X \ge x \} \le \frac{\langle X \rangle}{x},
    \label{eq:Markov_inequality_1}
\end{equation}
where the angled brackets denote the average over the possible realisations of $X$
and $ \Prob \{X \ge x \} $ is the complementary cumulative probability distribution of $X$.
A more general version of the bound~\eqref{eq:Markov_inequality_0}
valid for non-negative and non-decreasing functions $h$ of $X$, which can now have negative support,
is called the~\textit{Markov inequality} and reads
\begin{equation}
    \int_{-\infty}^{\infty} \mathrm{d} u \, h(u) f(u) \ge h(x) \int_{x}^{\infty} \mathrm{d} u \, f(u) .
    \label{eq:Markov_inequality_2}
\end{equation}

We now consider the function $h(x) = \e^{s x} $ for $s>0$ and use it in equation~\eqref{eq:Markov_inequality_2}
to obtain
\begin{equation}
    \Prob \{X \ge x \} \le \e^{-sx} \langle \e^{s X} \rangle.
    \label{eq:chernoff_bound}
\end{equation}
This inequality, well known in the probability community,
is referred to as the \textit{Chernoff bound}~\cite{Stewart2009probability}.
In the next section we will use the Chernoff bound 
to compare request and service in a communication channel
after introducing the notion of EB.

\section{Effective bandwidth}
\label{sec:Effective_bandwidth}

\subsection{Finite time}
\label{sec:Finite_time}
\begin{figure}
  \includegraphics[width=0.45\textwidth]{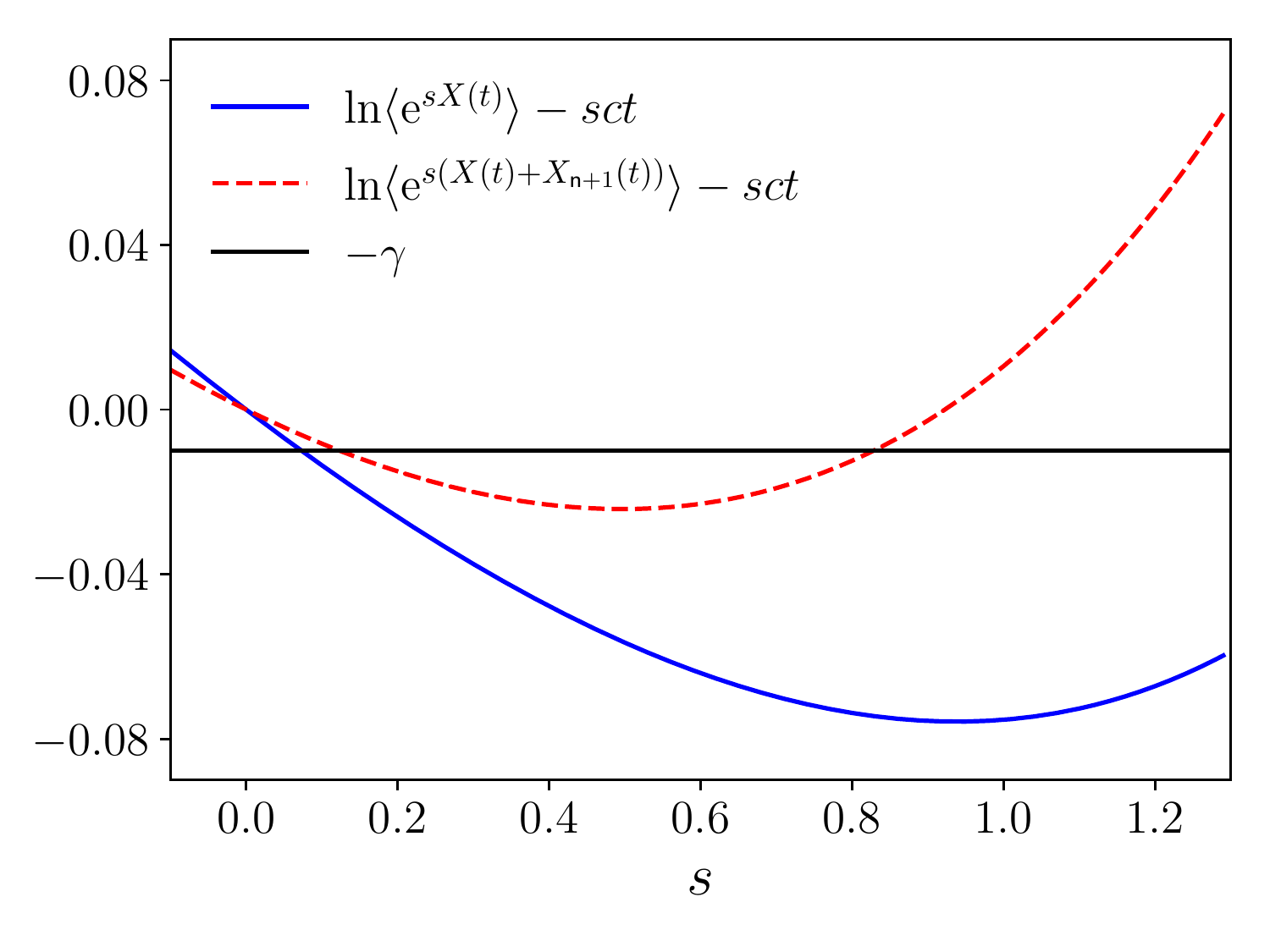}
  \caption
  {\label{fig:QoS}
  QoS control in finite time.  $X(t)$ and $X_{\mathsf{n}+1}(t)$ are Poisson processes of duration $t$
  with rates $\sum_{i=1}^{\mathsf{n}} \lambda_i =  0.09$ and $\lambda_{\mathsf{n}+1} = 0.05$.
  The QoS is set at $\gamma = -0.01$ , the capacity is $c=0.23$, and $t=1$.
  As there are values of $s$ such that the cumulant generating function of $X(t)+X_{\mathsf{n}+1}(t)$ is smaller than $s c t -\gamma$,
  the condition~\eqref{eq:QoS} for $X(t)+X_{\mathsf{n}+1}(t)$ is satisfied and the new request $X_{\mathsf{n}+1}(t)$ can be accepted.
}
\end{figure}

We define the \textit{finite-time EB function} $\Lambda(s,t)$ of a
process $X(t)$ of duration $t$, which describes a time-extensive
observable, as the functional
\begin{equation}
  \Lambda(s,t) = \frac{\ln \langle \e^{s X(t)} \rangle  }{s t},
  \label{eq:Finite_time_EF}
\end{equation}
where the angled-bracket notation here represents the average over histories of the process,
which are also referred to as \textit{trajectories}.
Following references~\cite{Kelly1996,Kelly2014}, we assume that a number $\mathsf{n}$
of sources contributes to an arrival process $X(t)$, i.e.,
\begin{equation}
    X(t) = \sum_{i=1}^{\mathsf{n}} X_i(t)  .
\end{equation}
According to the Chernoff bound~\eqref{eq:chernoff_bound},
the probability that the service  request  overflows the capacity satisfies
\begin{multline}
  \ln \Prob \{ X(t) > c  t \} \le \ln \langle \e^{s(X(t)-c  t)}\rangle\\
  = \ln  \langle \e^{s\,X(t)}\rangle - s   c  t,
\end{multline}
for all $s>0$. In this context, it is possible to introduce the notion of
\textit{quality of service} (for which the abbreviation QoS is commonly used
in engineering) in rigorous terms: for a given positive $\gamma$, we say that
QoS is guaranteed  if the condition
\begin{align}
  \Prob \{ X(t) > c  t\} \le \e^{-\gamma}
  \label{eq:QoS}
\end{align}
is satisfied.
For practical purposes, it is convenient to work with a stronger condition, i.e.,
\begin{equation}
    \inf_{s>0} \{ \ln \langle \e^{s X(t)}\rangle - s  c  t\} \le - \gamma,
\end{equation}
which is sufficient for equation~\eqref{eq:QoS} to hold. This means that if $
\ln  \langle \e^{s X(t)}\rangle - s  c  t$ is less than $-\gamma$ \textit{for
some} $s>0$, then the promise of QoS to the user is honoured. We are now in
a position to decide whether the server can accept another service request
$X_{\mathsf{n}+1}(t)$, from a source independent of $X(t)$, without violating
the condition~\eqref{eq:QoS}. The criterion is that the new request
$X_{\mathsf{n}+1}(t)$ is accepted if there is at least one value $s>0$ such that
\begin{equation}
   \ln  \langle \e^{s X(t)}\rangle + \ln \langle \e^{s X_{\mathsf{n}+1}(t)}\rangle - s  c  t \le -\gamma.
   \label{eq:QoS_2}
\end{equation}
Dividing the inequality \eqref{eq:QoS_2} by $s t$ yields
\begin{equation}
   \frac{\ln  \langle \e^{s X(t)}\rangle }{st}+ \frac{\ln \langle \e^{s X_{\mathsf{n}+1}(t)}\rangle}{s  t}  - c \le -\frac{\gamma}{s t},
   \label{eq:finite_time_EB}
\end{equation}
which  shows that the EBs naturally compare to the true bandwidth $c$. For
given values of $\gamma$ and $s$, the smaller the EB of a packet traffic model
is, the lesser the impact on the available resources  will be. A simple
example where the $X_i(t)$ are Poisson processes with rates $\lambda_i$,
$i=1,\ldots,\mathsf n+1$ is shown in figure~\ref{fig:QoS}, where it is easy to
check whether the arrival process $X_{\mathsf n+1}(t)$ can be accepted,
armed with the knowledge that the cumulant generating function for $X_i(t)$ is $\lambda_i t (\e^{s} -
1) $.

In short, we can now decide whether establishing a new connection can affect the
promised QoS for a given time by computing the finite-time EB functions of the incoming traffic
sources~\cite{Kelly2014}. Clearly, finite-time EB functions can also be calculated for the net
load and the available work.
Considering the long-time limit of $\Lambda(s,t)$ leads to natural connections to statistical mechanics
and to the cloning algorithm developed in that field, 
as discussed in the following subsections.

\subsection{Asymptotic analysis}
\label{sec:Asymptotic_analysis}
To study the long-time limit, we assume that the net-load process satisfies a large deviation
principle, loosely written in the form
\begin{equation}
  \Prob \{ W(t) =  w t\} \asymp \e^{-t \hat{e}(w) }
  \label{eq:large_deviation_principle}
\end{equation}
with rate function $\hat{e}(w)$, where the symbol $\asymp$ means logarithmic equality in the limit as $t \to \infty$.
The rate function $\hat{e}(w)$
encodes for the fluctuations around the typical value $\overline{w}$, which are of interest in resource allocation.
As an aside, if $\hat{e}(w)$ is convex, then the large deviation principle \eqref{eq:large_deviation_principle} implies
\begin{equation}
  \Prob \{ W(t) >  w t\} \asymp \e^{-t \hat{e}(w) }
  \label{eq:large_deviation_principle_2}
\end{equation}
for $w > \overline{w}$, as the probability on the left-hand side is dominated by the slowest decaying contribution.
The rate function can be obtained from the scaled cumulant generating function~(SCGF)
\begin{equation}
  e_W(s) = \lim_{t \to \infty} \frac{1}{t} \ln \langle \e ^{s W(t)} \rangle,
\end{equation}
when the latter is differentiable, by means of a Legendre--Fenchel (LF) transform
\begin{equation}
  \hat{e}(w) = \sup_s \{s w - e_W(s)\}.
  \label{eq:Legendre}
\end{equation}
The inverse of equation~\eqref{eq:Legendre} is also verified, i.e.,
\begin{equation}
    e_W(s) = \sup_w \{sw - \hat{e}(w)\},
    \label{eq:inverse_Legendre}
\end{equation}
as reviewed in reference~\cite{Touchette2009}.

More care is needed when the SCGF is non-differentiable. In fact, more generally, the LF transform of $e_W(s)$ yields the convex envelope of $ \hat{e}(w)$
which can contain linear sections corresponding to the non-differentiable points of $e_W(s)$,
interpreted as dynamical phase transitions~\cite{Bertini2006}.
In this paper, we will not deal with such circumstances.

Similarly to section \ref{sec:Finite_time}, we now turn our attention to the
event that  the service request overflows the capacity. Specifically, we
require that the net load exceeds a specified $q>0$ at any finite time $\tau$ in $(0,t]$.  To study this probability we can make the following heuristic argument based on the discrete-time analogue (see,
e.g., references \cite{Lewis1996, Srikant2013}).

First we note that to find the supremum of $W(\tau)$ it suffices to consider only the transition times $t_i$ and the final time $t$.  Hence we have the inequality
\begin{widetext}
\begin{equation}
  \Prob \left\{ \sup_{0 < \tau \le t} \, W(\tau) > q \right\} \leq \sum_{n=1}^\infty \int_0^t \mathrm{d}t_1 \int_0^t \mathrm{d}t_2 \ldots \int_0^t \mathrm{d}t_n \, p(t_1,t_2,\ldots,t_n) \sum_{i=1}^n \Prob \{ W(t_i) > q \} + \Prob \{W(t)>q \},
  \label{eq:net load_overflow}
\end{equation}
\end{widetext}
where $p(t_1,t_2,\ldots,t_n)$ is the joint probability density to have transitions at times $t_1, t_2,\ldots,t_n$.  We expect this inequality to become tighter for larger $q$ where the probability of exceedance at more than one time becomes small. 

Now, since the net load typically decreases in time (becomes more negative) according to the
stability condition of section~\ref{sec:Net_load}, the probability that $W(t)$
exceeds $q$ approaches zero as $t \to \infty$. Indeed we anticipate that the
most likely time for exceedance scales with $q$. For large $q$ and $t$ this suggests using~\eqref{eq:large_deviation_principle_2} to approximate $\Prob \{ W(t_i) > q \}$ by
$\e^{-q\frac{\hat{e}(q/t_i)}{q/t_i} }$. The right-hand side of~\eqref{eq:net
load_overflow}  is then dominated by the smallest exponent $\inf_{w \ge q/t} {\hat{e}(w)}/{w}$, where the condition $w \ge
q/t$ follows from $t_i \le t$.  For $t\gg q$, this infimum occurs at a
$t$-independent value $ \arg \inf \hat{e}(w)/w = q/\tau^*$, consistent with
our intuition that the most significant contribution to the probability \eqref{eq:net
load_overflow} comes from a time which scales with~$q$.

Putting everything together, we obtain
\begin{equation}
  \Prob \left\{ \sup_{0 < \tau \le t } \, W(\tau) > q\right\} \asymp \e^{- q \delta },
  \label{eq:exponential_decay_w}
\end{equation}
which expresses an important asymptotic result
for $t \to \infty$ and $q \to \infty$, with $q/t \to 0$.
The exponent $\delta$ is the largest number $s$ such that $s - \hat{e}(w)/w \le 0$ for all $w \ge 0$,
or the largest number $s$ such that $e_W(s)$ is non-positive, using equation~\eqref{eq:inverse_Legendre}.
Hence, the inequality
\begin{equation}
  e_W(\delta) \le 0
\end{equation}
is satisfied.

It is worth noting that the same arguments can be applied to study the asymptotic probability that the
queue occupation $Q(t)$ (rather than the net load) overflows $q$, which is
obviously more interesting for applications. In fact, from equation
\eqref{eq:queue}, we get
\begin{equation}
    \Prob \left\{ Q(t) > q \right\} = 
    \Prob \left\{ \sup_{0<\tau \le t} \int_{\tau}^{t} \mathrm{d} u \,\mathrm{w}(u) > q \right\},
\end{equation}
which, by the same heuristic argument used to obtain equations~\eqref{eq:net load_overflow} and \eqref{eq:exponential_decay_w}, gives
\begin{equation}
  \Prob \left\{ Q(t) > q \right\} \asymp  \e^{- q \delta }.
\end{equation}

As in section~\ref{sec:Finite_time}, we focus on
net loads of the form \eqref{eq:net load_form},
whose exponent $\delta$ is given by
\begin{equation}
  \delta(c) = \sup \{ s: e_X(s)/s \le c\},
  \label{eq:exponent}
\end{equation}
with  the assumption that the stability condition $\bar x < c$ is satisfied.
The functional
\begin{equation}
  \Lambda(s) = \frac{e_X(s)}{s}
  \label{eq:EF}
\end{equation}
represents the observable to be compared against
the true capacity $c$ and is referred to as
the \textit{asymptotic EB function} or simply the \textit{EB function}
of the process $X(t)$.
Note, we use the notation $\delta(c)$ in equation~\eqref{eq:exponent} 
and elsewhere when  we wish to emphasize dependence of the exponent on $c$.
The definition~\eqref{eq:EF} holds for generic extensive processes. However,
here we are chiefly interested in the case where $X(t)$ is the arrival process
and cannot decrease in time. This implies that its rate function has non-negative
support and the Legendre duality in turn implies that $\Lambda(s)$ is
a monotonic non-decreasing function of $s$.

As we did with equation~\eqref{eq:finite_time_EB}, we can assess the impact of
an incoming service request  on the available resources by computing the
residual $\Lambda(s)-c$. The stability condition ensures that, on average, all
the incoming requests are served, i.e.,  equation \eqref{eq:QoS} is satisfied
in the long-time limit for all real $\gamma$. On the other hand, it is of little
practical interest to consider the case where  $c$ is larger than the peak arrival rate, as
this would ensure an excellent QoS for all times, but at the cost of having
unused resources for most of the time. To find a more refined bound, let us
request that the probability decays faster than a certain reference law, i.e.,
\begin{equation}
  \Prob \left\{ \sup_{0 < \tau \le t} \, W(\tau) > q \right\} < \e^{-q  \xi } .
  \label{eq:ref_law}
\end{equation}
The exponent $\xi$ is a way of defining a target asymptotic QoS, and, obviously, the larger its value
the better service the final user is provided with. Using the
asymptotic relation~\eqref{eq:exponential_decay_w} 
yields $\delta(c) > \xi$, which implies
\begin{equation}
  \Lambda (\delta(c)) > \Lambda (\xi),
\end{equation}
due to monotonicity. 
We also have that $c = \Lambda (\delta(c))$, i.e., $\Lambda$ and $\delta$ are inverse functions.
This gives us the criterion
\begin{equation}
  c > \Lambda(\xi)
  \label{eq:criterion}
\end{equation}
to assess if the capacity $c$ is adequate to service the arrival process $X(t)$, given a target value of $\xi$.

It is worth noting that the notions that we have defined so far have glorious
analogues in equilibrium statistical mechanics~\cite{Reichl2009}, which we outline
here without aiming at being exhaustive.
In fact, equilibrium statistical mechanics is formulated in the
limit of the size $\mathfrak{n}$ of a macroscopic system approaching infinity, under the assumption that
the density $\Omega(\epsilon)$ of   microscopic states having a
mean energy $\epsilon$ satisfies
\begin{equation}
  \Omega(\epsilon) \asymp \e^{\mathfrak{n} \mathrm{s}(\epsilon)};
\end{equation}
the total energy $\mathfrak{n} \epsilon$, which is
extensive in the size, has the same role as the net load, which is extensive in
time, while  $\mathrm{s}(\epsilon)$ is the micro-canonical entropy function of the system and
is analogous to $\hat{e}_W(w)$.
Another fundamental quantity in statistical mechanics is the Helmholtz 
free energy $\mathrm{A}(\beta)$, which is obtained from the entropy 
function via an LF transform, with variable $\beta$ 
conjugated to the mean energy $\epsilon$, followed by scaling by $\beta$. 
Here $\beta$ is the reciprocal $1/(k_B T)$ of the system temperature $T$ 
and $k_B$ is the Boltzmann constant.
The function $\mathrm{A}(\beta)$ conveys the same information as the entropy
$\mathrm{s}(\epsilon)$ (when that function is strictly concave and differentiable),
but sometimes in thermodynamics it is more convenient to use the former than the latter.
In fact, the Helmholtz free energy is analogous to the EB function $\Lambda(s)-c$ for a net load of the form
\eqref{eq:net load_form}, where the parameter $s$ is conjugated to the time-averaged 
load. Similarly to the thermodynamic analogues, the EB function follows from
an LF transform and encodes for the same information as
$\hat{e}(w)$, while its use is recommended over the rate function if one wants to
assess to what extent  the fluctuations affect the~QoS.
We can control the value of the conjugated variable $s$ by setting it to
$\xi$, which represents the (rather arbitrary) target exponential tail of the
probability in equation~\eqref{eq:ref_law} and is positive.
The EB  $\Lambda(\xi)$ can be
thought of as a macroscopic quantity, which  describes the resource available
to the user   and can be derived from the microscopic dynamics of $X(t)$.
Obviously, $\Lambda(\xi)$ is additive in the number of service requests  as we
can obtain the EB of pooled independent processes by summing the individual
EBs of each process.

The focus on time-extensive observables $X(t)$, $Y(t)$, and $W(t)$
suggests even closer connections with \textit{non-equilibrium} statistical
mechanics, which essentially deals with time-extensive ``currents''
rather than with the size-extensive energy~\cite{Zia2007, Derrida2007, Touchette2013}.
As an aside, it is also worth noting that the asymptotic result~\eqref{eq:exponential_decay_w} 
is reminiscent of recent results on the universal statistics of extrema for
observables such as entropy~\cite{Neri2017}; see in
particular~\cite{Chetrite2018} for analysis of the housekeeping heat, which,
similarly to the net load, is an observable that on average decreases
with time.
The analogies outlined in this paragraph suggest the exploitation of methods borrowed from non-equilibrium
statistical mechanics,
such as the so-called ``cloning method'' for the
computation of $\Lambda(s)$. This is the topic of the next subsection.

\subsection{Monte Carlo evaluation}
\label{sec:Monte_Carlo_evaluation}

Computing large deviation functionals by means of Monte Carlo methods is hard,
as time-intensive observables  are doomed to converge to their typical values.
An approach that prevents such a fate from happening in simulations is the
``cloning method''~\cite{Anderson1975, DelMoral2004, Giardina2006, Giardina2011}.
This enables a desired functional to be evaluated directly by propagating an ensemble of trajectories
in time and cloning/pruning such trajectories appropriately. The cloning
method has deep roots in mathematical
physics~\cite{Metropolis1949,Anderson1975} and can be thought of as a sequential Monte Carlo (SMC) strategy,
also sometimes referred to as sequential importance resampling
(SIR)~\cite{Doucet2011}, tailored to sample large deviation events. While SMC
is typically presented in a discrete-time setting (but
see~\cite{Fearnhead2017} for rigorous details on SMC in continuous time), the
cloning method has been extensively used for continuous-time Markov processes,
as proposed in~\cite{Lecomte2007}. For the present EB formalism it is
convenient to adapt the procedure of reference~\cite{Cavallaro2016}, which
also includes the case of non-Markovian continuous-time processes. Central to
this procedure is to define and simulate the time $\tau$ between two
consecutive random events occurring at $t_{n}$  and $t_{n+1}$  and the
corresponding increment $\Theta_n$ in the value of the driven observable, as
defined for $n=1,2,\ldots$ in section \ref{sec:Net_load}; this can be easily
done for both fluid and point processes.
The aim is to estimate the SCGF
as the exponential growth of $\langle \e^{s X(t)}\rangle$,
which, loosely speaking, can be obtained from the ensemble of simulated trajectories
if each is cloned (or pruned) when its configuration changes from $n$ to $n + 1$
according to a factor $\e^{s\Theta_n}$,
while the average cloning (or pruning) rate is monitored.
More precisely, the algorithm consists of the following steps:

\begin{enumerate}[label=\arabic*)]
 \item Set up an ensemble of $N$ clones, each  with a time variable $t$, a random configuration $x_0$, and a counter $n=0$.
Set a variable $C$ to zero. For each clone, simulate the time $\tau$ until the next event and set $t$ to $\tau$.
Then, choose the clone with the smallest value of~$t$.
 \item \label{item:item} For the chosen clone increase the observable by an amount $\Theta_n$ and update $n$ to $n+1$.
 \item Increment the  value of $t$ for the chosen clone to $t+\tau$, where $\tau$ is the waiting time for the clone until the next event, i.e., until $t_{n+1}$.
\item 
 \label{item:cloning_step} Compute $y = \lfloor \e^{s \Theta_n}  + u\rfloor$, where $u$ is drawn from the uniform distribution on $[0,1)$.
\begin{enumerate}[label=\arabic*)]
  \item If $y=0$, prune the current clone. Then replace it with another one, uniformly chosen among the remaining $N-1$.
  \item If $y>0$, produce  $y$ copies of the current clone.
    Then, prune a number $y$ of elements, uniformly chosen among the existing $N+y$.
\end{enumerate}
\item Increment $C$ to $C + \ln [(N+\e^{s\Theta_n}-1)/N]$.
Choose the clone with the smallest $t$, and repeat from~\ref{item:item} until $t$ for a chosen clone reaches the desired simulation time $T$.
\end{enumerate}
The EB is finally recovered as $C/(s T)$ for large $T$. This prescription can suffer from finite-ensemble errors
 if any of the numbers $s\Theta_n$ is large enough for a single clone to replace  a conspicuous fraction of the existing ensemble elements. Such an effect can  be alleviated by choosing large~$N$; for further discussion on this and related points, see~\cite{Hurtado2009, Cavallaro2015, Nemoto2016, Brewer2018, Angeli2019}.

In the next sections we will consider selected Markovian and non-Markovian processes
(fluid processes in section \ref{sec:fluid_process} and point processes in sections \ref{sec:MMPP} and \ref{sec:Numerical_results_for_non-markovian_modulation})
and demonstrate that the cloning method reproduces exact analytical solutions within numerical accuracy
and thus can be reliably used as an automated way to compute the~EB.

\section{Markov fluid process}
\label{sec:fluid_process}
Introducing the EB in the previous section involved us defining a service
that is provided at a constant deterministic rate $c$. This led to the
notion of a fluid process, which describes the continuous flow in or out of a
source (or server)  subject to random periods of filling and emptying.
The fluid process was arguably first introduced by P.\ A.\ P.\ Moran to
describe the level of a dam, based on a discrete-time stochastic
process~\cite{Moran1954}. Since then,  continuous-time variants have also been
analysed, with remarkably many contributions to modelling of high-speed
data-networks, building on reference~\cite{Anick1982}.
Although Markov fluids are well known to many specialists in
traffic and queueing modelling~\cite{Gautam2012},
it is worth dedicating a whole section to them,
as they provide an elegant setting to demonstrate the EB and the cloning method.

We call an observable a \textit{Markov fluid} if its time derivative is a function of a
continuous-time Markov process. To illustrate this, we consider a server that can
be in many states, and during the stay in each state releases a fluid with
a certain deterministic rate. In such a model, deterministic and stochastic
dynamics coexist, and the traffic generated can be thought of as a piece-wise
continuous flow, in contrast with standard discrete-packet models. The flow
intensity varies according to the  state of an underlying continuous-time
stochastic process with state space $\mathcal S$ and generator $\mathbf
G$.~Each state $i \in \mathcal S$ corresponds to a different flow
intensity~$b_i$. This generates a process which we refer to as~$B(t)$ and can
generically represent an arrival, a service or a net-load process.

As an example of an arrival process we can combine  a number $\left|
\mathcal{S} \right|$ of identical sources, which can be active or inactive. In
this case the configuration state of the underlying Markov chain has dimension
$\left| \mathcal{S} \right|$, each state corresponding to the number of active
sources, with the total flow at its peak when all the sources are
active~\cite{Anick1982}.

For a general fluid, the rates $b_i$, $i=1,2, \ldots, \left| \mathcal S \right|$ are organised into
the matrix $\mathbf{B} = \mathrm{diag}(b_1,b_2,\ldots, b_{\left| \mathcal{S} \right|})$.
If the underlying Markov chain has transitions at times $t_1,t_2,\ldots, t_n$,
with $0 \le t_1 \le t_2 \le \ldots \le t_n \le t $,
the fluid process can be written as the \textit{continuous} functional
\begin{equation}
    B(t) = t_1 b_{\mathsf{s}(0)} + \sum_{i=1}^{n-1} (t_{i+1} - t_{i}) b_{\mathsf{s}(t_i)} + (t - t_{n}) b_{\mathsf{s}(t_n)},
\end{equation}
where $\mathsf{s}$ here simply maps the transition time to the underlying configuration
immediately after it.
We have chosen the symbol $B$ in allusion to the ``type B'' observables defined in
non-equilibrium statistical physics~\cite{Garrahan2009, Jack2015}. Similarly to such observables, each
term of $B(t)$ only depends on the configuration immediately after $t_i$ and the time
increment $t_{i+1} - t_{i}$ (except for the last term which only contributes
until $t$).

We now define a joint configuration--flow space and denote the probability
of having  a configuration $i$, at time $t$, with a value $\mathsf B$ for the 
observable $B(t)$ as $P_{i}(\mathsf B,t)$. This probability is propagated for a short time interval $\tau$, up to the first order in $\tau$, as
\begin{multline}
  P_{i}(\mathsf{B},t + \tau) =
    \sum_{j \neq i} [\mathbf{G}]_{ij} \tau  P_j(\mathsf{B},t) \\
    + (1 + [\mathbf{G}]_{ii} \tau ) P_i(\mathsf{B}-b_i \tau,t),
\end{multline}
which yields, in the limit as $\tau \to 0$, the master equation
\begin{equation}
    \frac{\mathrm{d}}{\mathrm{d}t} P_{i}(\mathsf B,t) = \sum_{j} [\mathbf{G}]_{ij} P_{j}(\mathsf B,t)
      - b_{i} \frac{\partial }{ \partial  \mathsf B}P_{i}(\mathsf B,t),
    \label{eq:label_B_type}
\end{equation}
where $i,j=1,2,\ldots, \lvert \mathcal S\rvert$ are generic configurations of the underlying Markov process and  $[\mathbf{G}]_{ij}$ is the generic entry
of $\mathbf{G}$. In matrix form we get
\begin{equation}
    \frac{\mathrm{d}  }{\mathrm{d}t}  | P(\mathsf B,t) \rangle = \mathbf{G}  | P(\mathsf B,t)  \rangle - \mathbf{B}  \frac{\partial }{\partial \mathsf B}  | P(\mathsf B,t)  \rangle ,
    \label{eq:Master_equation_fluid2}
\end{equation}
where $| P(\mathsf B,t) \rangle$  is the column  vector with entries
$P_1(\mathsf B,t), P_2(\mathsf B,t), \ldots, P_{\mathcal S}(\mathsf B,t)$. In
the limit as $t$ approaches infinity, the observable $B(t)/t$ concentrates
around its typical value
\begin{equation}
  \bar b = \langle 1 | \mathbf B | P^* \rangle,
  \label{eq:typical_bar_B}
\end{equation}
where $\langle 1 \rvert$ is the row vector with all entries equal to one and $| P^*
\rangle $ is the stationary distribution of the underlying Markov chain, i.e., the
solution of $\mathbf G | P^* \rangle= 0$.

Now, the full system can be diagonalised with respect to the subspace of the
$\mathsf{B}$s by means of an integral transform, which yields the biased
master equation
\begin{equation}
    \frac{\mathrm{d}}{\mathrm{d}t} \tilde{P}_{i}(t) = \sum_{j} [\mathbf{G}]_{ij}\tilde{P}_{j}(t)  +  s b_{i} \tilde{P}_{i}(t),
    \label{eq:biased_fluid}
\end{equation}
where $\tilde{P}_{i}(t) = \int_0^\infty  \mathrm{d} \mathsf{B}\, \e^{s \mathsf B} P_i (\mathsf B, t)$ and we assume that the boundary conditions are such that
 $\e^{s \mathsf B} P_i(\mathsf B,t)  \vert_0^{\infty} =0$;
more care is needed when these boundary conditions are not satisfied, but cloning simulations confirm that the approach works well in practice.
Equation~\eqref{eq:biased_fluid} corresponds to the dynamics of
a system that changes configuration according to $\mathbf{G}$, whose weight
evolves exponentially with rate $s b_{i}$ during the stay in state $i$, and
feeds $B(t)$ an amount $b_{i} \tau$ after each visit to $i$ of
duration~$\tau$.
In matrix form this is equivalent to
\begin{equation}
    \frac{\mathrm{d}  }{\mathrm{d}t}  | \tilde{P}(t) \rangle = (\mathbf{G} + s \mathbf{B} ) | \tilde{P}(t) \rangle
    \label{eq:Master_equation_fluid}
\end{equation}
and has formal solution
\begin{equation}
    | \tilde{P}(t) \rangle = \exp \left[(\mathbf{G} + s \mathbf{B})  t \right]  | \tilde{P}(0) \rangle.
    \label{eq:formal_solution_fluid}
\end{equation}

The finite-time EB of a Markov fluid can be expressed as follows. The aim is
to find $\Lambda(s,t)=\ln \langle \e^{s B(t)} \rangle /(s t)$, where the
angled-bracket expectation value is obtained by averaging over both the
configuration and flow space,~i.e.,
\begin{equation}
 \langle \e^{s B(t)} \rangle =   \int \mathrm{d} \mathsf{B} \, \langle 1| \e^{s \mathsf{B}}|P(\mathsf{B},t) \rangle .
\end{equation}
Integrating over $\mathsf{B}$ component-wise and using equation~\eqref{eq:formal_solution_fluid}
gives the form 
\begin{equation}
    \Lambda(s,t) = \frac{1}{s t}\ln \langle 1  |  \exp \left[ (\mathbf{G} + s \mathbf{B}) t \right] | \tilde{P}(0)\rangle ,
    \label{eq:Markov_fluid_effective_bandwidth}
\end{equation}
which also shows that the EB can be obtained by propagating in time
an initial state $| \tilde{P}(0)\rangle $.

As a simple example, we focus now on a source modulated by a
\textit{telegraph process}, i.e., a two-state ($1$ and $2$)
continuous-time Markov process with generator
\begin{equation}
   \mathbf{G} = \left( \begin{array}{cc}
    -\alpha & \beta  \\
     \alpha & -\beta \\
  \end{array} \right),
\label{eq:telegraph_generator}
\end{equation}
where $\alpha$ and $\beta$ are transition rates. The service request is produced deterministically at rate $b_1$ or $b_2$, when the configuration is $1$ or $2$, respectively,
while the typical behaviour is given by
\begin{equation}
  \overline{b} = \frac{\beta b_1 + \alpha b_2}{\alpha +\beta }.
  \label{eq:average_b}
\end{equation}
The biased master equation explicitly reads
\begin{equation}
    \frac{\mathrm d}{\mathrm dt}
    \left( \begin{array}{c}
        \tilde P_{1}(t)\\
        \tilde P_{2}(t)
    \end{array}\right)
    =
    \left( \begin{array}{cc}
    -\alpha + b_1 s & \beta  \\
     \alpha & -\beta + b_2 s \\
    \end{array} \right)
    \left( \begin{array}{c}
        \tilde P_{1}(t)\\
        \tilde P_{2}(t)
        \end{array} \right).
        \label{eq:fluid_biased_master}
\end{equation}
Such a model can describe a source that is either  in an idle state, i.e., transmitting only a
few packets, or in a active state and transmitting at its peak rate.
Further assuming that the observation starts in the stationary distribution
\begin{equation}
  | P^* \rangle  = \left( \begin{array}{c}
        \frac{\beta}{\beta+\alpha}\\
        \frac{\alpha}{\beta+\alpha}
        \end{array} \right),
\end{equation}
we can compute the finite-time EB function
\begin{equation}
  \Lambda(s,t) = \frac{1}{st} \ln   \langle1| \exp  \left[ \left( \begin{array}{cc}
    -\alpha + b_1 s & \beta  \\
     \alpha & -\beta + b_2 s \\
    \end{array} \right) t \right] \, | P^* \rangle .
\end{equation}
 
We now turn to consider the asymptotic properties of a stable net load, fed by a telegraph-modulated fluid source and serviced  by a channel of constant capacity $c$. Such a system
is described by a master equation equivalent to equation~\eqref{eq:label_B_type},
with the addition of a loss term $c \, \partial P_{i}(\mathsf B,t)/\partial
\mathsf  B$, which in matrix form reads
\begin{equation}
    \frac{\mathrm{d}  }{\mathrm{d}t}  | P(\mathsf B,t) \rangle = \mathbf{G}  | P(\mathsf B,t)  \rangle - (\mathbf{B} - c \mathds{1} ) \frac{\partial  }{\partial \mathsf B}  | P(\mathsf B,t)  \rangle.
    \label{eq:Master_equation_fluid1}
\end{equation}
The observable $B(t)$ now represents a net load of the form~\eqref{eq:net load_form}. 
Upon a bilateral integral transform, equation~\eqref{eq:Master_equation_fluid1} can be written instead as
\begin{equation}
    \frac{\mathrm{d}}{\mathrm{d}t}  | \tilde{P}(t) \rangle = \left[ \mathbf{G} + s (\mathbf{B} - c \mathds{1})\right] | \tilde{P}(t) \rangle,
\end{equation}
where the entries of $|\tilde P(t)\rangle$ are $\tilde{P}_{i}(t) =
\int_{-\infty}^\infty \mathrm{d} \mathsf{B} \,\e^{s \mathsf B} P_i (\mathsf B, t)$ and
we now assume that the boundary conditions are such that $\e^{s \mathsf B } P_i(\mathsf B,t) \vert_{-\infty}^{\infty} = 0$.

\begin{figure}
  \centering
  \includegraphics[width=0.45\textwidth]{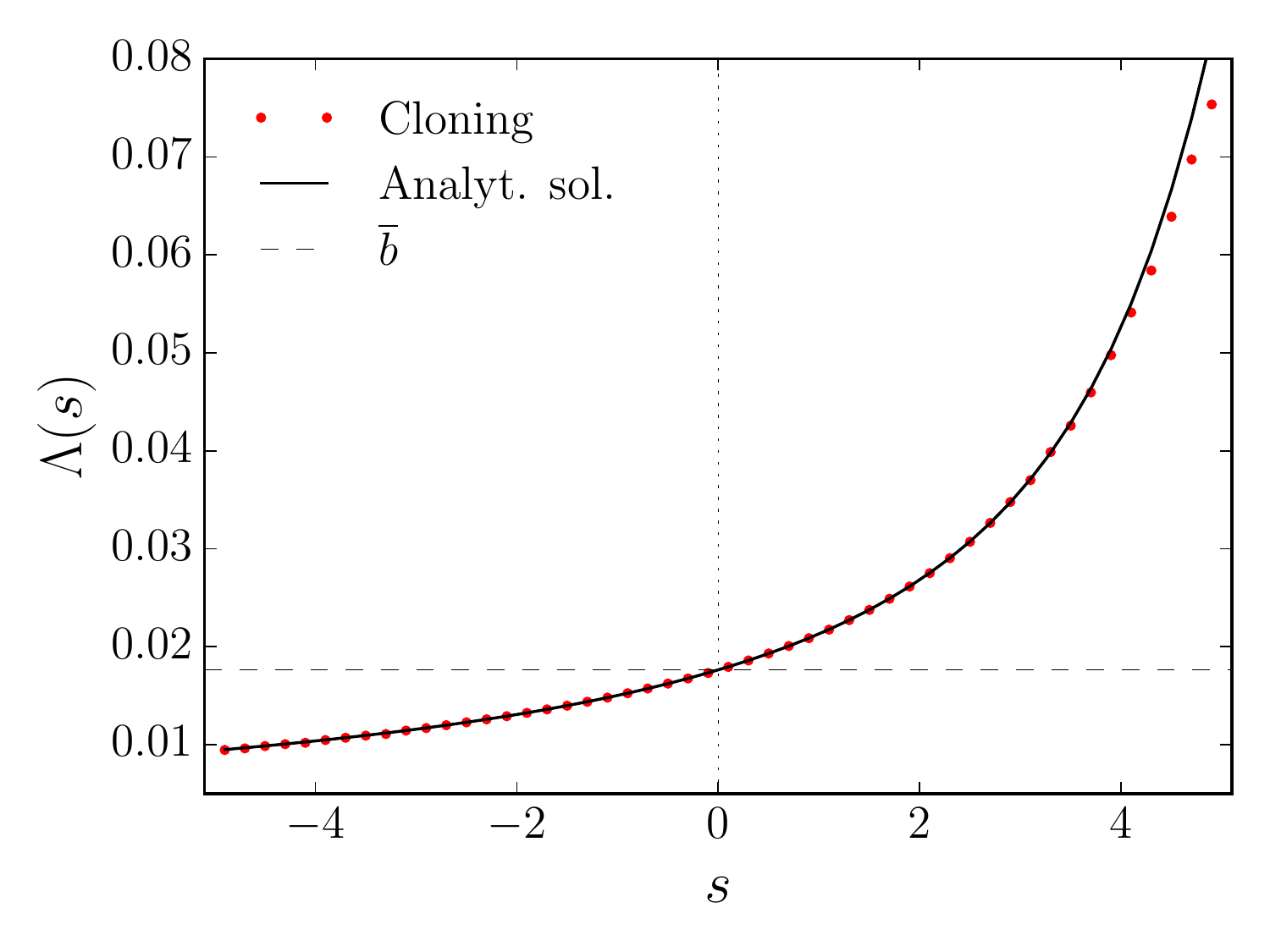}
  \caption{
    \label{fig:eb_vs_s_1}
    EB function for the telegraph fluid model with parameters $(\alpha, \beta, b_1, b_2)=(0.1,5,0,0.9)$.
    Markers correspond to cloning simulations with $(t,N)=(10^4,10^4)$, which underestimate the exact solution
    for large values of $s$ due to finite-ensemble effects (see main text).
  }
\end{figure}

For the fluid modulated by a telegraph process
it is straightforward to verify that the asymptotic EB function is given by
\begin{multline}
  \Lambda(s) =  - \frac{1}{2 s} \Big\{\alpha + \beta - (b_1 + b_2)s  \\
- \sqrt{[\alpha + \beta - (b_1+ b_2)s]^2 - 4[b_1 b_2 s^2  - (b_1 \beta + b_2 \alpha)s]}  \Big\} ,
\label{eq:Lambda_fluid}
\end{multline}
which is the leading eigenvalue of the biased generator of equation~\eqref{eq:fluid_biased_master},
divided by $s$.
In figure~\ref{fig:eb_vs_s_1} it is shown that the result
 obtained using the cloning method of subsection
\ref{sec:Monte_Carlo_evaluation} is consistent with the exact analytic solution, 
except for well documented finite-ensemble errors~\cite{Hurtado2009, Cavallaro2016} at large positive values of~$s$.
Now, we are ready to tackle the problem of deciding
whether, for this fluid traffic model,
it is possible to guarantee a promised asymptotic QoS in the terms of
inequality~\eqref{eq:ref_law} with target $\xi$.
The criterion \eqref{eq:criterion}
ensures that as long as the EB $\Lambda(\xi)$ is smaller than $c$,
the user receives  service as agreed with the provider.

\section{Markov modulated Poisson process}
\label{sec:MMPP}
In this section, we consider sources modelled by \textit{Markov modulated
Poisson processes} (MMPPs), which can be thought of as the discrete
counterparts of the fluid sources seen in the previous section,
and, similarly to those, are ubiquitous in traffic and queueing modelling~\cite{Fischer1993, Stewart2009probability, Gautam2012}.
As in fluid
processes, in MMPPs, the load is modulated by the generator $\mathbf{G}$ and
its intensity is defined by the diagonal matrix
$\mathbf{B}=\mathrm{diag}(b_1,\ldots,b_{\lvert \mathcal S\rvert})$.
When the
Markov chain is in state $i$, with $i=1,\ldots, \lvert \mathcal S\rvert$, events
occur according to a pure birth process (Poisson process) with state space
$\mathbb{N}^{0}$ and rate $b_i$.
In fact, MMPPs and fluid processes share a
number of additional features; for example, the parallelism between the
concept of EB function for fluid processes and MMPPs is detailed in
reference~\cite{Elwalid1993}. Similarly to fluid processes, MMPPs have been
extensively used for  telecommunication modelling~\cite{Fischer1993} (see also the application in appendix~\ref{app2})
 but have
also  been exploited for biological modelling~\cite{Dobrzynski2009, Cavallaro2019}. A further
interesting remark is that, even if MMPPs are modulated by a Markovian
stochastic generator, the sequence of arrival events is non-Markovian, hence
this framework can be used, in general, to model events that occur with
time-varying rates~\cite{Fischer1993}. If increments occur at times
$t_1,t_2,\ldots, t_n$, the count process can be written as the functional
\begin{equation}
    B(t) = \sum_{i=1}^{\infty} \mathbbm{1}_{t_i<t},
\end{equation}
where $\mathbbm{1}_x$ is the indicator function of $x$.

\begin{figure}
  \centering
  \includegraphics[width=0.45\textwidth]{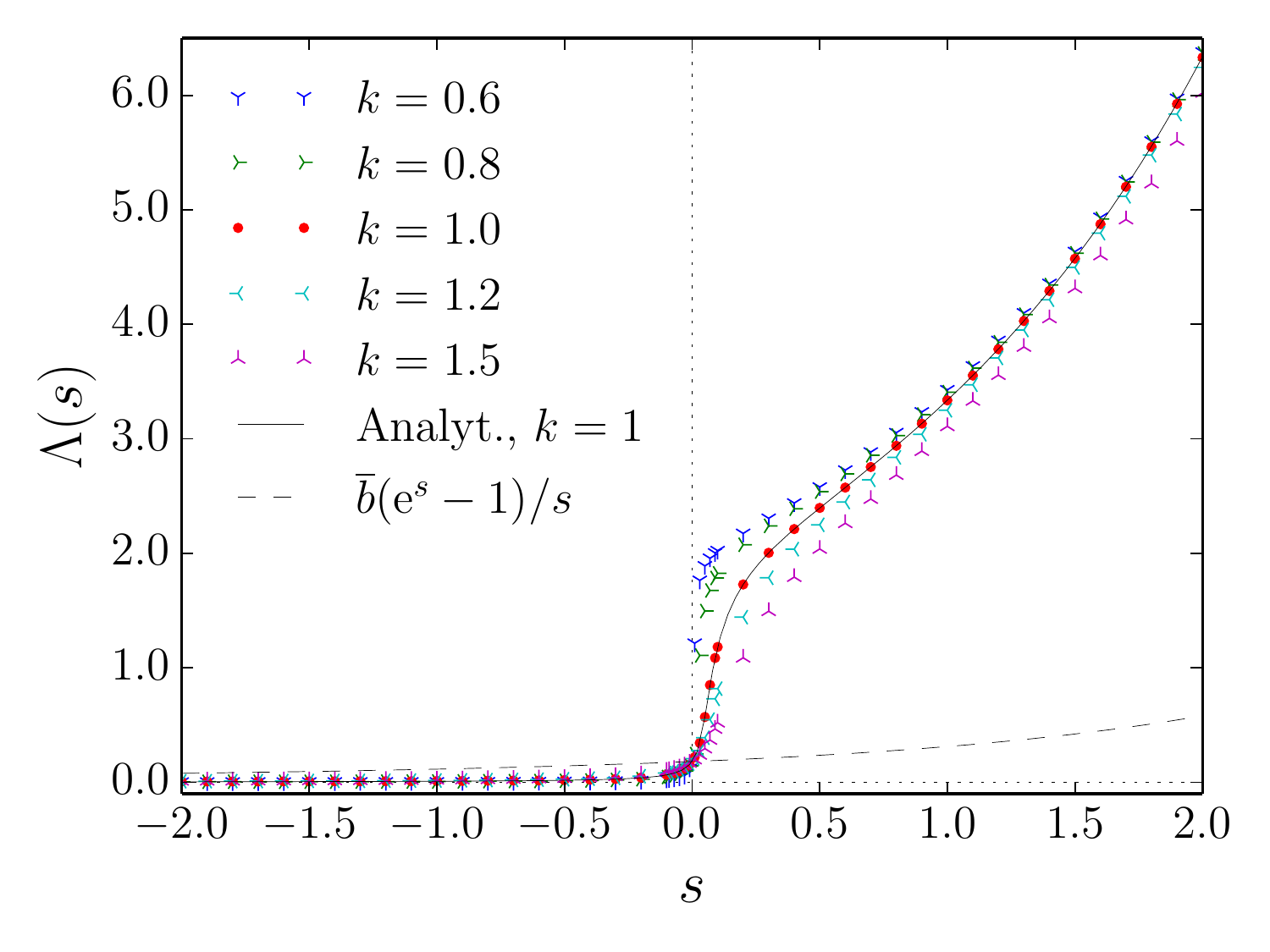}
  \caption{
  \label{fig:eb_vs_s}
  EB function for the two-state modulated Poisson process with $(b_1, b_2) =
  (0, 2)$ and Weibull distributed switching times \eqref{eq:weibull0} and
  \eqref{eq:weibull1}. Weibull distributions have rate parameters as defined
  in \eqref{eq:rate_weibull}, with $(\alpha, \beta) = (0.01, 0.1)$ and varying
  values of the shape parameter $k$. The case $k = 1$ corresponds to an MMPP.
  Markers are results of cloning simulations with $(t, N) = (10^4,10^4)$. For
  comparison, the dashed line indicates the EB of a homogeneous Poisson
  process of rate $\bar{b}$.}
\end{figure}

The master equation for the joint MMPP and its underlying Markov process  is
\begin{equation}
  \frac{\mathrm d}{ \mathrm d t} P_{i,\mathsf B}(t) = \sum_j [\mathbf G]_{ij} P_{j,\mathsf B}(t) + b_i P_{i,\mathsf B-1}(t) - b_i P_{i,\mathsf B}(t),
  \label{eq:joint_MMPP}
\end{equation}
where $\mathsf B$ is the number of birth events which have occurred by time $t$ and $i,j = 1,2,\ldots, \lvert \mathcal S \rvert$
are generic configurations of the modulating Markov process.
A matrix representation akin to equation~\eqref{eq:Master_equation_fluid1} is
\begin{equation}
    \frac{\mathrm{d}  }{\mathrm{d}t}  | P(t) \rangle =\mathbf{G} \otimes \mathds{1}| P(t) \rangle + \mathbf{B} \otimes (a^+ - \mathds{1}) 
    | P(t)  \rangle ,
\end{equation}
where $| P(t) \rangle \in \mathcal S \otimes \mathbb{N}^{0}$
has elements $P_{i,\mathsf B}(t)$, while $a^+$ and
$\mathds{1}$ are the creation and the identity operators, respectively, in
$\mathbb{N}^{0}$. After a  transform with $\tilde{P}_{i}(t) =
\sum_{\mathsf B} \e^{s\mathsf B} P_{i, \mathsf B}(t) $ and 
suitable boundary conditions,
equation~\eqref{eq:joint_MMPP} can be written as
\begin{equation}
  \frac{\mathrm d}{ \mathrm d t} \tilde{P}_{i}(t) = \sum_j [\mathbf G]_{ij} \tilde{P}_{j}(t) + b_i (\e^s-1)\tilde{P}_{i}(t).
\end{equation}
In vector form we have
\begin{equation}
  \frac{\mathrm{d}}{ \mathrm{d} t} | \tilde{P}(t) \rangle = [\mathbf{G}   +  \mathbf{B}(\e^s-1) ] | \tilde{P}(t) \rangle,
  \label{eq:MMPP_vector_form}
\end{equation} 
where $ | \tilde{P}(t) \rangle$ has entries $\tilde P_1 (t), \tilde P_2 (t),\ldots, \tilde P_n (t)$.
The finite-time EB function can be formally expressed as
\begin{equation}
 \Lambda(s,t) = \frac{1}{st} \ln \langle 1 | \exp\{ [\mathbf G + (\e^s -1 ) \mathbf B] t \} | \tilde{P} (0) \rangle ,
\end{equation}
which is an analogue of equation~\eqref{eq:Markov_fluid_effective_bandwidth}.
As a simple toy model, we consider the Poisson process modulated by the
telegraph process with generator \eqref{eq:telegraph_generator}. Despite its
simplicity, such a model captures the stochastic dynamics of gene expression
when the gene is not self-regulated and switches between high and low activity
phases, see, e.g., references \cite{Kepler2001, Horowitz2017, Tiberi2018}. Its
EB function  can be easily derived as
\begin{widetext}
\begin{equation}
\Lambda(s) = - \frac{1}{2 s}
\Big\{ \alpha + \beta - (b_1 + b_2) (\mathrm{e}^{s}-1)
 - \sqrt{ [\alpha + \beta - (b_1 + b_2) (\mathrm{e}^{s}-1) ]^2 - 4 [b_1 b_2 (\mathrm{e}^{s}-1)^2 - (b_1 \beta + b_2 \alpha ) (\mathrm{e}^{s}-1) ] }  \Big\}.
\label{eq:Lambda_MMPP}
\end{equation}
\end{widetext}
The random variable $B(t)/t$ concentrates around the most probable value
$\overline{b}$ given by equation~\eqref{eq:average_b}. It is also worth noting
the similarity in the form of equation~\eqref{eq:Lambda_MMPP} with the fluid EB function of equation~\eqref{eq:Lambda_fluid};
indeed the two expressions are the same to first-order approximation in $s$.
 In figure~\ref{fig:eb_vs_s} the MMPP EB function
(solid line in the figure) is shown to be accurately reproduced by cloning
simulations (dot markers).

As a remark, in the modulated Poisson processes, there are several net-load
increments between two configuration changes, each increment being of unit
magnitude. As a result, finite-ensemble effects are smaller than those of the
fluid net load, where, in our simulations, increments are added only at
configuration changes and depend exponentially on the inter-event time.

\section{Non-Markov modulated Poisson process}
\label{sec:Numerical_results_for_non-markovian_modulation}

Finally, and significantly, we introduce a generalization of the two-phase model described in
section~\ref{sec:MMPP} by now considering that the arrival rate of the Poisson
process is modulated by a semi-Markov process.
Here the distributions of the lengths of the phases 1 and 2
are non-exponential thus exacerbating temporal correlations.
Our choice is to draw the duration of phases 1 and 2 from the Weibull distributions
\begin{align}
  f_1(t|k, \tilde \alpha) &= k  \tilde \alpha^k t^{k-1} \mathrm e^{-(\tilde \alpha t)^k}, \label{eq:weibull0}\\
  f_2(t|k, \tilde \beta) &=  k  \tilde \beta^k t^{k-1} \mathrm e^{-(\tilde \beta t)^k} \label{eq:weibull1},
\end{align}
with shape $k$ and rates $\tilde \alpha$ and $\tilde \beta$, respectively.
The rate parameters are chosen to be
\begin{equation}
  \tilde \alpha =  \alpha \, \Gamma(1 + 1 / k), \qquad 
  \tilde \beta = \beta \, \Gamma(1 + 1 / k),
  \label{eq:rate_weibull}
\end{equation}
where $\Gamma(x)$ is the Gamma function.
This guarantees that each phase has the same mean duration as the process considered in
the previous section and that the resulting modified arrival process converges to
the same typical value~\eqref{eq:average_b}; tuning the parameter $k$ only alters the fluctuation scenario.

Generally, for a non-Markov model such as the one described in this section, analytical progress is difficult; however, the
cloning method remains a powerful way to evaluate the EB and thus assess
the QoS numerically.
This is illustrated in figure~\ref{fig:eb_vs_s}, where the EB functions for our semi-Markov modulated processes
(non-dot markers in the figure)  are compared to the standard MMPP case (which is
obtained for $k=1$). The figure also shows the analytical result for a (homogeneous)
Poisson process with identical value of $\overline{b}$, which clearly has very
low effective bandwidth---adding a source of this type to the
net load has little effect on the QoS.

On increasing the value of $k$  in the semi-Markov modulated process, the distributions of the durations become more
peaked around their expected values, the traffic can be thought of
as being more regular, and the effective bandwidth decreases (recall that $s>0$ is the relevant case here). Similarly,
broad phase-length distributions (obtained by decreasing the value
of $k$) correspond to strong fluctuations and  high EB.

\begin{figure}
  \centering
  \includegraphics[width=0.45\textwidth]{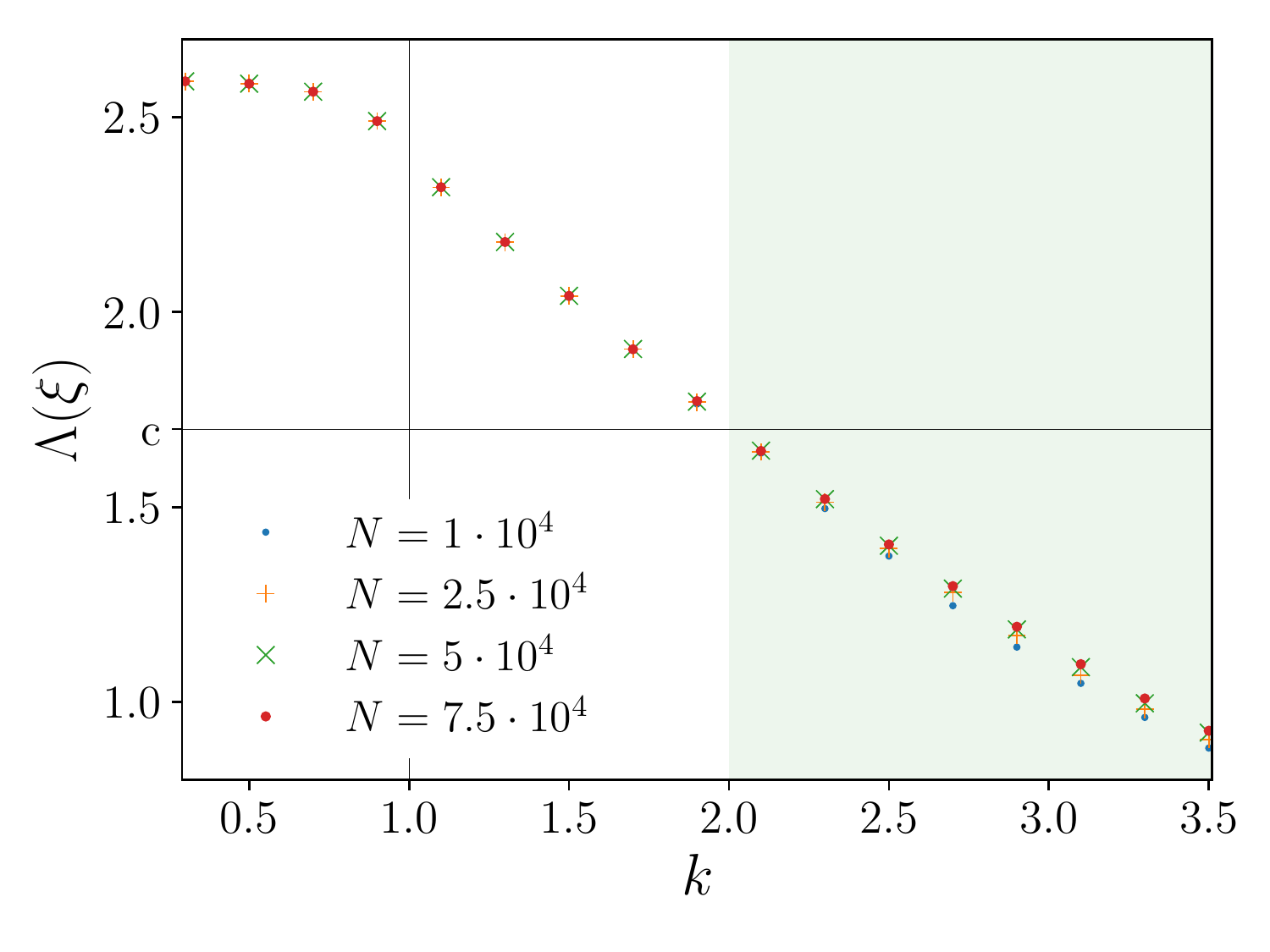}
  \caption{\label{fig:eb_vs_k}
    Cloning-simulation EB of semi-Markov modulated Poisson process with target exponent and bandwidth $(\xi,c)=(0.5,1.7)$.
    Parameters $\alpha, \beta, t$ are as in figure~\ref{fig:eb_vs_s}.
    Simulations with different values of $N$
    are compared and suggest that 
    finite-ensemble effects do not play a significant role, at least \mbox{for~$k<2$}.}
\end{figure}

As a demonstration, we target an exponential decay with specific exponent $\xi$ for the
net load and plot the EB $\Lambda(\xi)$ as function of $k$ in
figure~\ref{fig:eb_vs_k}. For a given service capacity $c$, the statistics of
the arrival requests can be
modified by changing $k$ so that $\Lambda(\xi) < c$ (shaded area in figure~\ref{fig:eb_vs_k}) in order
to provide the desired asymptotic QoS whilst
maintaining the mean arrival rate~\eqref{eq:typical_bar_B}. This type of
modification is akin to the so-called ``traffic
shaping''~\cite{MischaSchwartz1996}.
As alluded to above, having more regular on and off periods lowers the EB,
thus leaving resources available for other requests but still maintaining the mean
traffic $\bar b$.
We see in figure~\ref{fig:eb_vs_k} that, for $c=1.7$, this simple toy model
achieves a target exponent of $\xi=0.5$ when $k$ exceeds a value around~$2$.
Arguably, similar considerations apply to more realistic models 
whose EBs can also be investigated by cloning;
having a general numerical method to obtain $\Lambda(s)$ finally allows the use of
criterion \eqref{eq:criterion} to assess the asymptotic QoS.

\section{Discussion}
\label{sec:discussion}

While physicists have been regarding the large deviation theory as an elegant
way to formulate statistical mechanics, teletraffic engineers and operational
researchers have been using large deviation results to estimate the likelihood
that a demand in service overflows the available resources.  A central role in
teletraffic engineering is played by the effective bandwidth (EB) function
$\Lambda(s)$. This function facilitates the construction of a neat criterion to
decide whether a promised ``quality of service'' can be maintained  in finite
time and in the long-time limit, despite the threat of disruptive rare events.
In this paper, we reviewed the concept of EB, showing that the function
$\Lambda(s)$ can be thought of as a Helmholtz free energy and demonstrating
that the cloning method, which has been developed in non-equilibrium
statistical physics, is a general numerical scheme for the evaluation of
$\Lambda(s)$.

The notion of EB is based on comparing the incoming requests (forming the
packet traffic) with a very idealised protocol to process them, i.e.,
deterministic service of constant rate $c$. This is easily demonstrated in the Markov fluid process,
for which  analytical and numerical
solutions are shown to be in excellent agreement. Nevertheless, non-Markovian
systems are among those that can benefit the most from tools to predict the
probability of large deviation events as has been discussed, e.g., in
recent physics literature~\cite{Andrieux2008,Maes2009,Cavallaro2015,
Harris2015, Cavallaro2016, Sughiyama2018}. Hence, we focused on models
from two classes of  non-Markovian arrival processes (viz., the Markov and semi-Markov
modulated Poisson processes) showing  the extent to which the service-request
statistics affect the EB. Simulation results on simple on-off toy models show
that having more regular on and off periods lowers the EB, thus leaving
resources available for other requests.

One can conceive further augmentation of the standard Markov modulated Poisson
process (MMPP) by replacing, e.g., the modulated Poisson process with a
generalised birth-death process. This generalization has been used to model
populations in randomly switching environments modulated by a Markov generator
(see, e.g.,~\cite{Hufton2016} and references therein).
The lowest-order approximation of modulated birth-death processes leads to
the \textit{piecewise-deterministic Markov processes}
(PDMPs)~\cite{davis1984piecewise}, which have also been recently shown to be
appropriate for the natural sciences (where the underlying Markov process
represents the extrinsic noise)~\cite{Zeiser2010,Realpe-Gomez2013,
Hufton2016,Lin2016}. Such PDMPs are reminiscent of the Markov fluid of
section~\ref{sec:fluid_process}, but more general than that, as the
deterministic evolution between jumps can be
non-linear and the transition rate of the underlying process can depend on the
 fluid variable.
We believe that the cloning method and the EB can contribute to
straightforwardly exploring these more complex processes upon defining the
correct cloning factor, with
exponent given by an increment $\Theta_n$ which is non-linear in $\tau$.
It is also worth mentioning that a non-equilibrium statistical mechanics of
PDMPs has been presented in references~\cite{Crivellari2010,Faggionato2009}.

Despite the fact that the specialised literature is rich in exact solutions
(mainly for Markov processes, see,
e.g.,~\cite{bucklew1990large,shwartz1995large,Weiss1995,Kelly1996,Lewis1996}
and references therein), a systematic and general numerical scheme  to
compute the EB function $\Lambda(s)$ may be of practical interest.  In this
contribution we have argued that the cloning method of non-equilibrium
statistical mechanics provides such a scheme and, significantly, can also be
applied to non-Markovian processes. In fact realistic traffic models often
incorporate memory as they convey the patterns of human dynamics, which are
non-Markovian~\cite{Huberman1997, Barabasi2005}. Hence, there
is potentially much more that could be done in terms of actual applications,
such as  in the  validation of traffic
protocols and beyond.

\begin{acknowledgments}
 Most of this research was performed while MC was a PhD student at
 Queen Mary University of London.  We warmly thank Ra\'ul J.\ Mondrag\'on
 for bringing the problem of performance evaluation in telecommunications
 to our attention  and for sharing valuable insights.
 RJH gratefully acknowledges an External Fellowship from the London Mathematical Laboratory.
 The research utilised Queen Mary's MidPlus computational facilities,
 supported by QMUL Research-IT and funded by EPSRC grant EP/K000128/1.
\end{acknowledgments}

\appendix

\section{Statistics of packet loss}
\label{app2}

Let us consider  modelling the occupation of a queue by a one-dimensional random walk on a linear chain of length
$\mathsf{N}$. When the walker is in position $\mathsf N$, a new arrival (with
rate $\lambda$) causes the total-current counter to tick, but leaving the
occupation number (i.e., the underlying  configuration $i$) unchanged. Such a
system has a lucid interpretation in queuing theory and is referred to as an
M/M/1/\textsf{N} queue in the so-called \textit{Kendall} notation~\cite{Stewart2009probability}. Customers
arrive according to a Poisson process at rate $\lambda$ and are processed by a
single server at rate $\mu$,  but there is space in the server for only $\mathsf{N}$ customers. When the server is fully occupied, there is no
interruption of the arrival process; however the new customers do not alter
the queue, simply disappearing instead. In communication systems, such
customers are said to be ``lost''.  Formally, the occupation number of the
queue follows a birth-death process, where the new arrivals can be neglected
when $i \ge \mathsf{N}$. Hence, the stationary state has grand-canonical
distribution
\begin{equation}
  P^*_i =   \frac{(1-\lambda/\mu)}{\left[1-(\lambda/\mu)^{\mathsf{N}+1}\right]}   (\lambda/ \mu)^i, \quad i \le \mathsf{N},
\end{equation}
if $\lambda \neq \mu$, or $P^*_i = 1/(\mathsf{N}+1)$, if $\lambda = \mu$, and
satisfies the detailed-balance condition. Now suppose we are interested in the statistics
of particle loss, i.e., we want to count the number of customers that arrive
when the occupation number of the queue is $\mathsf N$. In the context of the present paper, we point out that this can be encoded into an MMPP with $[\mathbf B]_{ii}=0$, $i<\mathsf N$, and  $[\mathbf
B]_{\mathsf N \mathsf N}=\lambda$. The mean packet-loss rate is simply given by the
arrival rate $\lambda$ times the probability  $P^*_\mathsf{N}$ that the queue
is full. Such arrivals correspond to jumps that leave that state as it is, but still
contribute a factor $\e^{s}$ in the modified
dynamics (defined in equation~\eqref{eq:MMPP_vector_form}).
The MMPP framework thus provides one way to access fluctuations of the packet loss around its mean rate.

\bibliographystyle{plain}
\bibliography{mybib}

\end{document}